\documentclass[referee]{aa}
\usepackage{graphicx}
\usepackage{txfonts}
\begin{document}
   \title{Physical parameters of T dwarfs derived from high-resolution near-infrared spectra}
   \subtitle{}

   \author{C. del Burgo
          \inst{1}
          \and
          E. L. Mart\'\i n\inst{2,3}
          \and
          M. R. Zapatero Osorio \inst{2}
          \and
          P. H. Hauschildt \inst{4}
          }

   \offprints{C. del Burgo}

   \institute{School of Cosmic Physics. Dublin Institute for Advanced Studies, Dublin 2, Ireland\\
              \email{cburgo@cp.dias.ie}
         \and
              Instituto de Astrof\'\i sica de Canarias, E-38200, La Laguna, Tenerife, Spain\\
              \email{ege@iac.es, mosorio@iac.es}
         \and
              University of Central Florida, Physics Department, PO Box 162385, Orlando, FL32816, USA\\
              \email{ege@physics.ucf.edu}
         \and
              Hamburger Sternwarte, Gojenbergsweg 112, D-21029 Hamburg, Germany\\  
              \email{yeti@hs.uni-hamburg.de}
             }

   \date{Received XXX XX, 2008; accepted XXX XX, 2009}

% \abstract{}{}{}{}{} 
% 5 {} token are mandatory
 
  \abstract
   {}
   {We determine the effective temperature, surface gravity and 
    projected rotational velocity of nine T dwarfs from the 
    comparison of high-resolution near-infrared spectra and 
    synthetic models, and estimate the mass and age of the 
    objects from state-of-the-art models.
    }
   {We use the AMES-COND cloudless solar metallicity models provided 
    by the PHOENIX code to match the spectra of nine T-type field 
    dwarfs observed with the near-infrared high-resolution 
    spectrograph NIRSPEC using ten echelle orders to cover part of
    the J band from 1.165 to 1.323 $\mu$m with a resolving power 
    $R\sim$20,000. The projected rotational velocity, effective 
    temperature and surface gravity of the objects are determined 
    based on the minimum root mean square of the differences between 
    the modelled and observed relative fluxes. Estimates of the 
    mass and age of the objects are obtained from effective 
    temperature-surface gravity diagrams, where our results
    are compared with existing solar metallicity models.}
   {The modelled spectra reproduce quite well the observed features 
    for most of the T dwarfs, with effective temperatures in the 
    range of 922-1009\,K, and surface gravities between $10^{4.3}$ 
    and $10^{5.0}$ $cm s^{-2}$. Our results support the assumption 
    of a dust free atmosphere for T dwarfs later than T5, where dust 
    grains form and then gravitationally sediment into the low 
    atmosphere. The modelled spectra do not accurately mimic some
    individual very strong lines like the K\,{\sc i} doublet at 1.2436 and 
    1.2525 $\mu$m. Our modelled spectra does not match well the observed 
    spectra of the two T dwarfs with earlier spectral types, namely 
    SDSSp J125453.90-012247.4 (T2) and 2MASS J05591914-1404488 
    (T4.5), which is likely due to the presence of condensate 
    clouds that are not incorporated in the models used here. 
    By comparing our results and their uncertainties to evolutionary 
    models, we estimate masses in the interval $\approx$5--75 $M_J$ 
    for T dwarfs later than T5, which are in good agreement 
    with those found in the literature. We found apparent young ages 
    that are typically between 0.1 and a few Gyr for the same T dwarfs, 
    which is consistent with recent kinematical studies.}
   {}

   \maketitle
%
%________________________________________________________________

\section{Introduction}

The first near-infrared spectrum of a T dwarf (Gl 229B) showed a 
predominance of methane (CH$_4$) bands that made it look similar to that 
of Jupiter (Oppenheimer et al. 1995) despite of a difference in temperature 
of about 800\,K. Atmospheric models were soon developed to fit the spectrum 
of Gl 229B to estimate its surface gravity, effective temperature, 
age and mass (Allard et al. 1996; Marley et al. 1996; Tsuji et al. 1996).
These models indicated that the atmosphere of this brown dwarf was free from 
dust grains, which were needed to explain the properties of the 
warmer L dwarfs (Allard et al. 1997). The dust grains in Gl 229B could be 
condensed in clouds dominated by organic compounds (Fegley et al. 1996; 
Griffith et al. 1998). It has also been proposed that a warm dust layer 
could be present deep inside the photosphere (Tsuji et al. 1999) and that 
the alkali resonance lines have very strong pressure-broadened red wings 
that provide a significant source of opacity at near-infrared wavelengths 
(Burrows et al. 2000). 
 
The discovery of free-floating objects similar to Gl 229B (Burgasser 
et al. 1999; Cuby et al. 1999; Strauss et al. 1999) indicated that these 
objects are numerous in the solar vicinity. A unified near-infrared 
classification scheme for T dwarfs has been defined by Burgasser et al. (2006). 
Theoretical models provide a good correspondence to the broad-band colors and 
low-resolution spectra of T dwarfs (Burrows et al. 2006). Currently, the 
coolest T dwarfs known have estimated effective temperatures between 600\,K 
and 700\,K (Warren et al. 2007; Delorme et al. 2008).  

Most modelling efforts on T dwarfs have concentrated on fitting broad-band 
colors and low-resolution spectra, and deriving atmospheric parameters
by comparing the data with synthetic spectra (Cushing et al. 2008,
Leggett et al. 2007, Saumon et al. 2007, Saumon et al. 2006,
Burgasser et al. 2006, Tsuji et al. 2005, Tsuji et al. 2004,
Burgasser et al. 2004). Mart\'\i n \& Zapatero Osorio (2003) 
estimated the surface gravity and effective temperature of one T dwarf from 
a mid-resolution near-infrared spectrum. High-resolution (R$\sim$20,000) 
observations in the $J$-band obtained with NIRSPEC on the Keck II telescope 
are now available (Zapatero Osorio et al. 2006, Mc Lean et al. 2007). 

Modelling the atmospheres and spectra of T dwarfs presents a set of challenges.
In the {\em AMES-COND} models used in this work (Allard et al. 2001), 
the profile of individual strong IR lines are approximated by 
Voigt profiles with estimated damping constants as no better
data are currently available for these lines. The line databases for
molecules such as FeH, CaH, CH$_4$ are not complete and in many cases lack
accuracy. However, in general the molecular line data are quite good as data
sources such as HITRAN are of high quality. The lower the effective
temperatures are, the better the completeness of the molecular line data as the
(comparatively) less accurate higher energy levels are less populated and thus
have smaller effects at lower temperatures.

In the {\em AMES-COND} models a dust-free configuration is used,
where the dust particles form but do not contribute to the opacities 
(Allard et al. 2001).
This assumption should be reasonable for cool T dwarfs, but will become
progressively worse for effective temperatures above $\approx$1200\,K.
Overall, the systematic errors are still considerably larger than,
e.g., for solar type stars, but the situation can only improve if 
more comparison between models and data are made to identify the 
areas where improvements are most urgently needed.

In this paper, we present the first comparison with {\em AMES-COND} models 
of high-resolution near-infrared spectra of a sample of nine T dwarfs 
observed by Zapatero Osorio (2006) that are used to derive their 
physical parameters. A summary of the observations used here is 
given in Section \ref{obsmoddata}. Section \ref{analysis} is 
devoted to the analysis of the data. The results are presented
in Section \ref{results}. Section \ref{discussion} contains the discussion
of our results, and finally, Section \ref{conclusions} contains our conclusions.

%__________________________________________________________________

\section{Observations, models and data preparation}
\label{obsmoddata}

\subsection{Observations}
\label{observations}

High-resolution near-infrared spectra in the $J$-band of nine T dwarfs
were obtained using the Keck\,II telescope and the NIRSPEC spectrograph. 
The instrumental setup was chosen to provide a wavelength coverage from 
1.148 up to 1.346\,$\mu$m split into ten different orders, a nominal dispersion 
ranging from 0.164 (blue wavelengths) to 0.191~\AA\,pix$^{-1}$ (red wavelengths), 
and a final resolution element of 0.55-0.70~\AA\ at 1.2485\,$\mu$m (roughly 
the central wavelength of the spectra), corresponding to a resolving power 
$R=17,800-22,700$. A detailed description of the observations and data reduction 
is provided in Zapatero Osorio et al. (2006) (hereinafter ZO06), who presented 
the T dwarfs that we used in this study.

In our spectroscopic analysis we have used as many echelle orders as possible: 
from order 57 up to order 66; this depends on the quality (in terms 
of signal-to-noise ratio) of the observed spectra. Table~2 in McLean et al. 
(2007) summarizes the wavelength and dispersion properties of each of the 
$J$-band NIRSPEC echelle orders 58 to 65 for the same instrumental configuration 
that we have used here.
Order 57 covers the wavelength interval 1.327--1.347\,$\mu$m with 
a nominal dispersion of 0.191~\AA\,pix$^{-1}$. Order 66 covers the range 
1.147--1.164\,$\mu$m with a nominal dispersion of 0.167~\AA\,pix$^{-1}$.

   \begin{figure}
   \centering
   \includegraphics[width=9.cm]{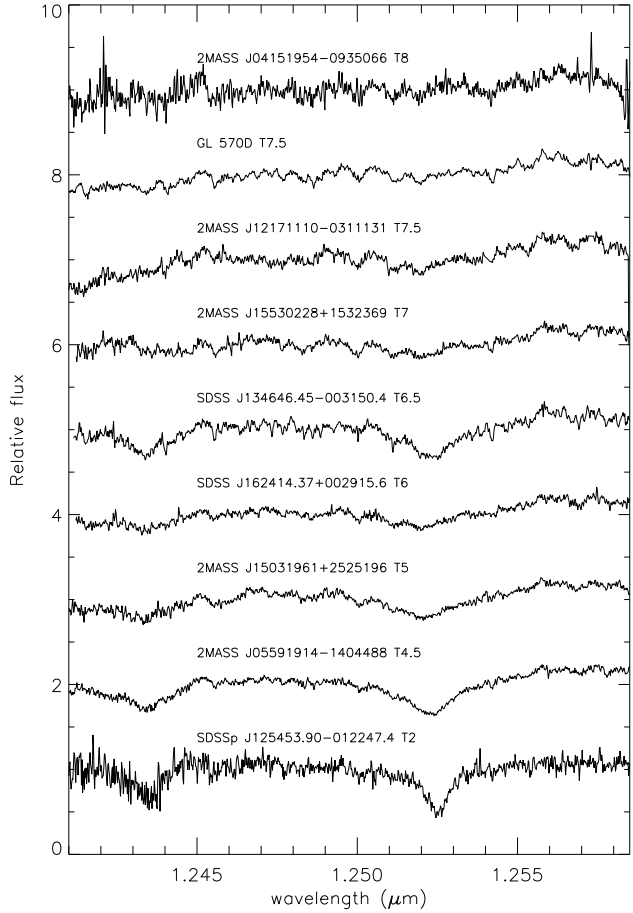}
\caption{Variation of the strong K\,{\sc i} doublet at 1.2436 and 1.2525 $\mu$m
         along the spectral type sequence of T dwarfs. All spectra are 
         normalized to unity over the entire wavelength range 
         and offset by a constant. Also, each spectrum
         is shifted in velocity to vacuum wavelengths.}
              \label{FigCdB1}%
   \end{figure}

Fig. \ref{FigCdB1} shows, as an example, the order 61 with the presence of 
the K\,{\sc i} doublet at 1.2436 and 1.2525 $\mu$m, for all the T dwarfs.

\subsection{Description of atmosphere models}
\label{models}

The PHOENIX code (Hauschildt \& Baron 1999) is a general purpose
stellar atmosphere modelling package that includes very complex atomic 
models and line blanketing by hundreds of millions of atomic and molecular 
lines. The code can be used to compute model
atmospheres and  synthetic spectra of cool objects. 
The radiative transfer in PHOENIX is solved in spherical geometry using
an operator splitting method (Hauschildt 1992, 1993).
At the low temperatures of the brown dwarfs,
a rich chemistry in the atmospheres of these objects is present, with 
hundreds of gas-phase species, liquids and crystals, and the formation 
of tens of different types of dust grains (e.g., silicates, amorphous 
carbon, iron). There are four different scenarios for the dust formation 
considered in PHOENIX, among which is the {\em AMES-COND} cloudless models 
(Allard et al. 2001).

Here we used a grid of the {\em AMES-COND} v2.2 models and synthetic spectra 
of solar metallicity, with effective temperatures $T_{\rm eff}$ ranging from 
700 to 3000 \,K (steps of 100\,K) and surface gravities $log\,g$ ranging 
from 3.0 to 5.5 (steps of 0.5) with $g$ in $cm\,s^{-2}$, to mimic the 
observed spectra of T dwarfs.

The {\em AMES-COND} v2.2 models use the H$_2$O line list of
Barber et al. (2006), the FeH list of Dulick et al. (2003), and
the damping constants used for the K \,{\sc i} lines are published 
in Allard et al. (2003).

\begin{figure}
   \centering
   \includegraphics[width=9.25cm]{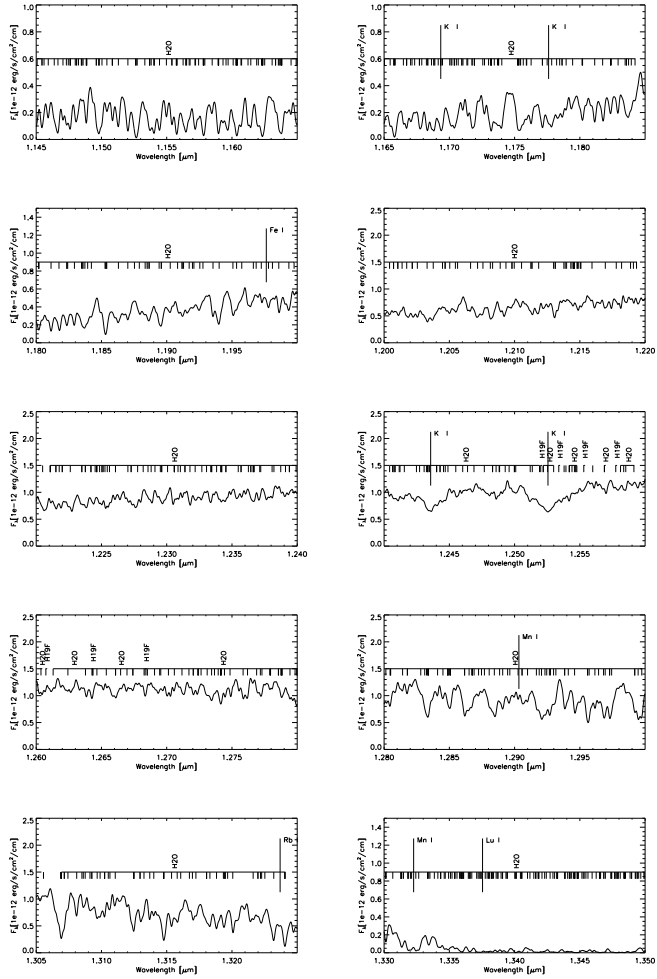}
\caption{Synthetic flux versus wavelength. 
         Identifications of some features of a synthetic spectrum
         with $T_{\rm eff}$=1000\,K and $log\,g$ = 4.5. Most of the lines
         are due to water vapor.}
              \label{FigCdB2}%
   \end{figure}

Fig. \ref{FigCdB2} illustrates the identification of some
near-infrared features over a synthetic spectrum with $v_{\rm
rot}$\,sin$i$\,=\,0, $T_{\rm eff}$\,=\,1000\,K and log\,$g$\,=\,4.5.
Most features are due to water vapor, and the only visible atomic 
lines are due to the K\,{\sc i} doublet at 1.25 $\mu$m.

\begin{figure*}
   \centering
   \includegraphics[width=14.5cm]{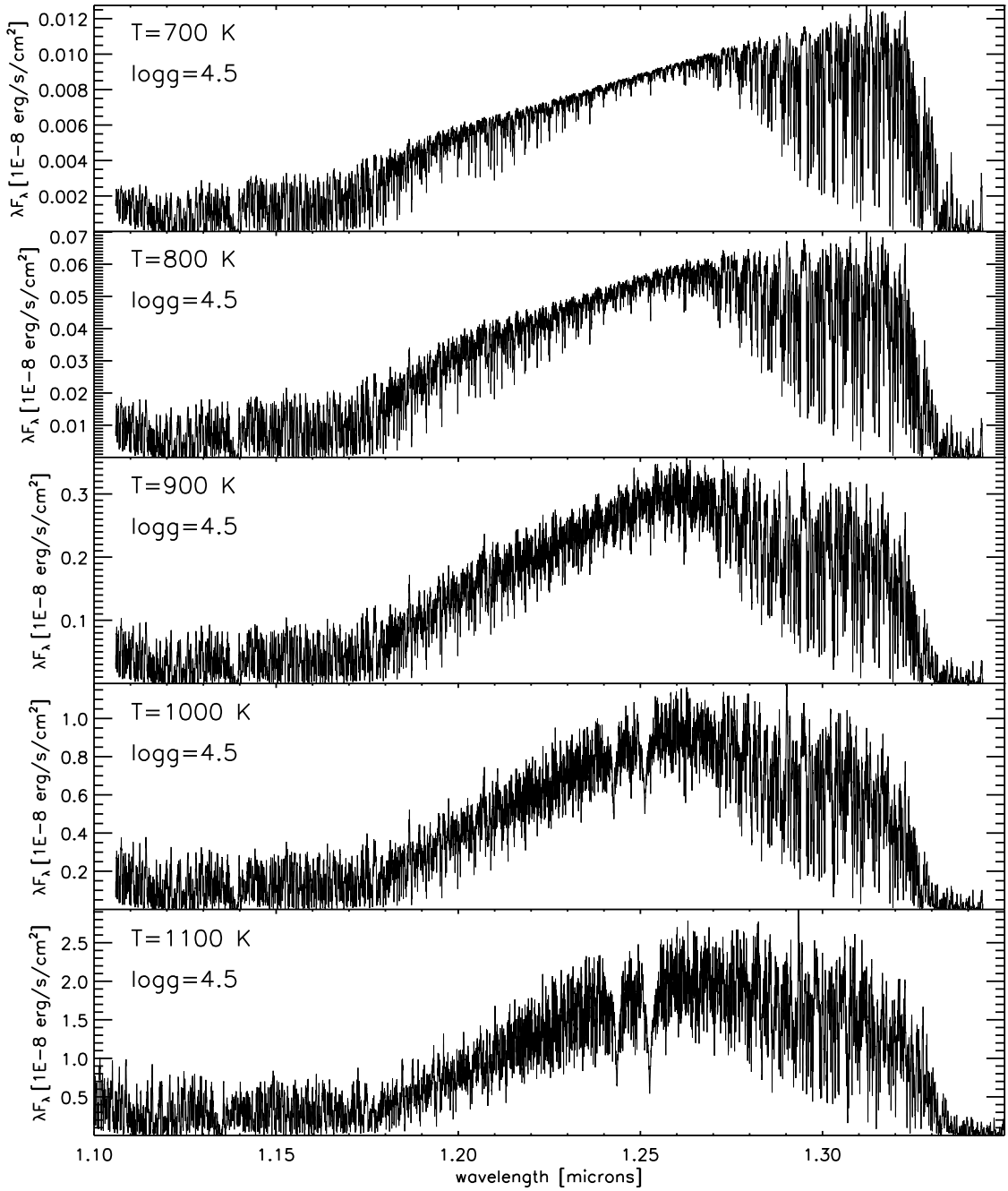}
\caption{Synthetic flux versus wavelength. Models with $log\,g$ = 4.5 (CGS units) and $T_{\rm eff}$=700, 800, 900, 1000 and 1100\,K. The synthetic spectra were computed with 0.05 \AA pix$^{-1}$, but are shown after being smoothed with a boxcar average of 10 pixel width.}
              \label{FigCdB3}%
   \end{figure*}

\begin{figure*}
   \centering
   \includegraphics[width=14.5cm]{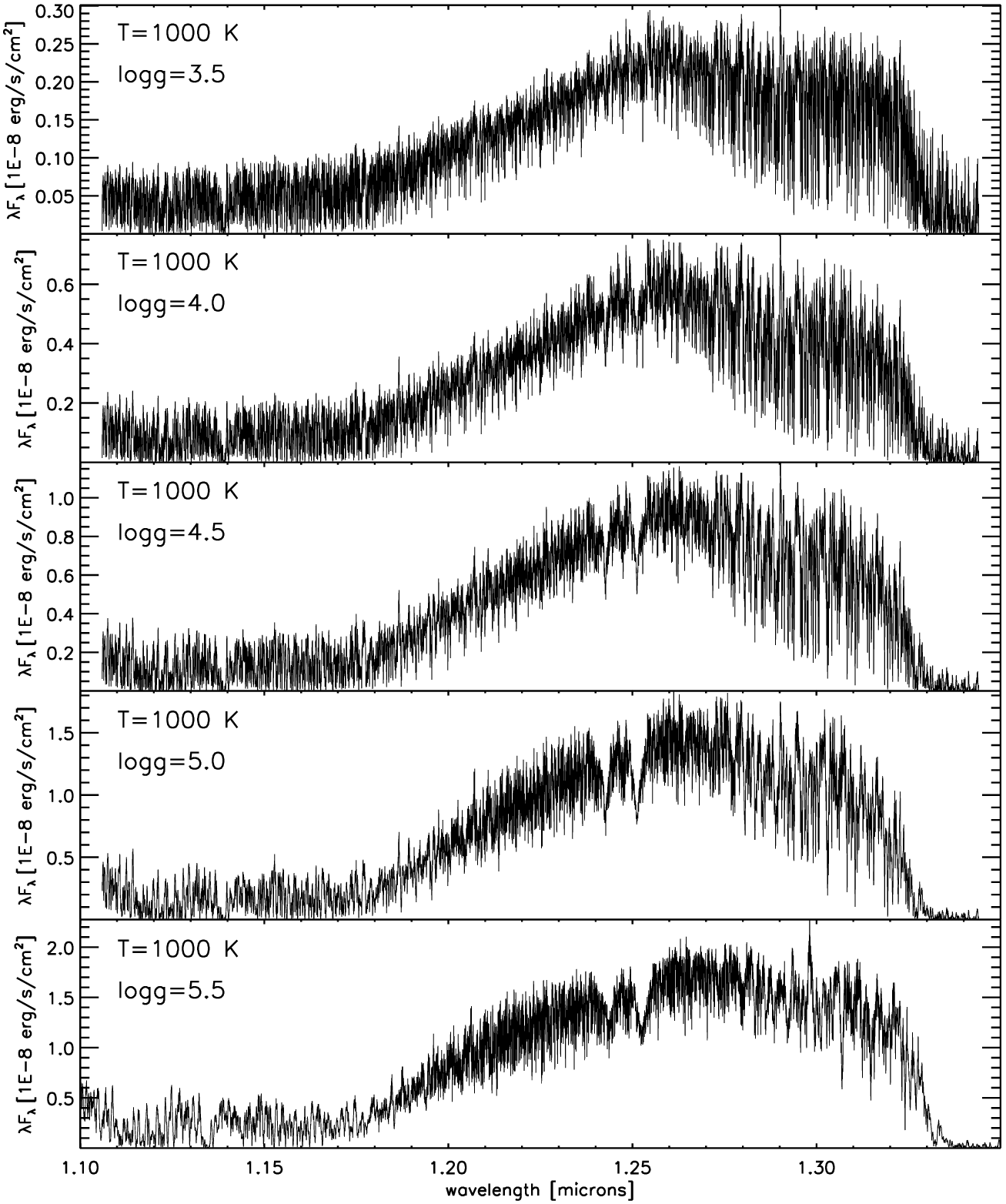}
\caption{Synthetic flux versus wavelength. Models with $T_{\rm eff}$=1000\,K and $log\,g$ = 3.5, 4.0, 4.5, 5.0 and 5.5 (CGS units). The synthetic spectra were computed with 0.05 \AA pix$^{-1}$, but are shown after being smoothed with a boxcar average of 10 pixel width.}
              \label{FigCdB4}%
   \end{figure*}

Fig. \ref{FigCdB3} shows the variations of the synthetic spectra for a
surface gravity of $log\,g$ = 4.5 and different effective temperatures
from 700 to 1100\,K. We note the strong change of the K\,{\sc i} doublet 
at 1.2436 and 1.2525 $\mu$m with temperature. Fig. \ref{FigCdB4} 
shows the variations of the synthetic spectra for various $log\,g$ 
(from 3.5 to 5.5) and the same $T_{\rm eff}$=1000\,K.

\subsection{Data preparation}
\label{dataprep}

All of our observed spectra were moved to vacuum wavelengths 
(i.e., laboratory frame of reference) for a proper comparison with the 
theoretical models. This was done from a cross-correlation analysis.
In a first iteration, we used the radial velocities published in Zapatero 
Osorio et al. (2007). Then, we determined the atmospheric parameters for
each T dwarf as explained later and used the corresponding synthetic spectra
as templates to recompute the radial velocities. New atmospheric parameters 
were thus obtained consistently with the recalculated radial velocities. We
note that our radial velocities are in agreement, within the errors, with 
those of Zapatero Osorio et al. (2007).

The grid of PHOENIX synthetic spectra were modified in order to be
compared with the NIRSPEC observations.
First, the synthetic spectra were transformed to take into account 
the projected rotational velocity 
($v_{\rm rot}$\ sin$i$) of the objects using the formalism of Gray (1992), 
with a limb darkening parameter $\epsilon$=0.6. A grid of models
with projected rotational velocities between 10 and 50 Km s$^{-1}$, with
steps of 1 Km s$^{-1}$, were generated. These spectra were also convolved 
with a Gaussian that mimics the instrumental profile along the 
dispersion axis for each order. The resulting modelled spectra were 
rebinned to the same resolution of the observations. Modelled spectra 
are normalized over the wavelength range corresponding to order 59.
All these steps were performed using our own programmes.

\begin{table}
\caption{SDSSp J125453.90-012247.4. Synthetic and observed spectra comparison 
         for all orders: RMS, N, $T_{\rm eff}$, $log\,g$ and $v_{\rm rot}$\ sin$i$.}             
\label{table:1}      
\centering          
\begin{tabular}{lccccc}     % 41 columns 
\hline\hline       
 order & N    & RMS    & $T_{\rm eff}$ & $log\,g$ & $v_{\rm rot}$\ sin$i$ \\
\hline  
 66    &  996 & 0.2797 & 1600 & 5.5 & 38 \\
 65    &  995 & 0.1740 & 1300 & 5.5 & 46 \\
 64    & 1001 & 0.1336 & 2400 & 3.5 & 36 \\
 63    & 1002 & 0.0931 & 2500 & 3.5 & 26 \\
 62    & 1002 & 0.0839 & 2700 & 3.5 & 27 \\
 61    & 1002 & 0.0960 & 2700 & 5.5 & 46 \\
 60    &  890 & 0.0914 &  700 & 5.5 & 28 \\
 59    & 1002 & 0.0811 & 2000 & 3.0 & 36 \\
 58    & 1002 & 0.0932 & 1500 & 5.5 & 36 \\
 57    &  994 & 0.0862 & 2000 & 4.0 & 31 \\
\hline                  
\end{tabular}
\end{table}

\begin{table}
\caption{2MASS J05591914-1404488. Synthetic and observed spectra comparison 
         for all orders: RMS, N, $T_{\rm eff}$, $log\,g$ and $v_{\rm rot}$\ sin$i$.}             
\label{table:2}      
\centering          
\begin{tabular}{lccccc}     % 41 columns 
\hline\hline       
 order &  N   & RMS   & $T_{\rm eff}$ & $log\,g$ & $v_{\rm rot}$\ sin$i$ \\
\hline  
 66    &  987 & 0.1079 & 1500 & 5.0 & 21 \\
 65    &  989 & 0.0796 & 1200 & 5.5 & 25 \\
 64    &  991 & 0.0575 & 1000 & 3.5 & 33 \\
 63    &  991 & 0.0512 &  800 & 5.5 & 24 \\
 62    &  991 & 0.0416 &  800 & 5.5 & 27 \\
 61    &  991 & 0.0612 & 1000 & 5.0 & 33 \\
 60    &  882 & 0.0434 &  800 & 5.5 & 25 \\
 59    &  991 & 0.0724 &  900 & 3.5 & 33 \\
 58    &  991 & 0.0682 & 1300 & 5.5 & 33 \\
 57    &  957 & 0.0491 & 1500 & 4.0 & 33 \\
\hline                  
\end{tabular}
\end{table}

\begin{table}
\caption{2MASS J15031961+2525196. Synthetic and observed spectra comparison 
         for all orders: RMS, N, $T_{\rm eff}$, $log\,g$ and $v_{\rm rot}$\ sin$i$.}             
\label{table:3}      
\centering          
\begin{tabular}{lccccc}     % 41 columns 
\hline\hline       
 order & N    & RMS & $T_{\rm eff}$ & $log\,g$ & $v_{\rm rot}$\ sin$i$ \\
\hline  
 66    &  969 & 0.1112 & 1300 & 5.0 & 34 \\
 65    &  978 & 0.0627 & 1100 & 4.0 & 42 \\
 64    &  987 & 0.0542 &  900 & 4.0 & 30 \\
 63    &  987 & 0.0445 & 1000 & 3.5 & 42 \\
 62    &  987 & 0.0413 &  800 & 5.5 & 37 \\
 61    &  987 & 0.0576 & 1000 & 5.5 & 42 \\
 60    &  873 & 0.0461 &  800 & 5.5 & 31 \\
 59    &  987 & 0.0619 &  900 & 3.5 & 37 \\
 58    &  987 & 0.0633 & 1300 & 5.0 & 33 \\
 57    &  911 & 0.0448 & 1400 & 4.5 & 33 \\
\hline                  
\end{tabular}
\end{table}

\begin{table}
\caption{SDSS J162414.37+002915.6. Synthetic and observed spectra comparison 
         for all orders: RMS, N, $T_{\rm eff}$, $log\,g$ and $v_{\rm rot}$\ sin$i$.}             
\label{table:4}      
\centering          
\begin{tabular}{lccccc}     % 41 columns 
\hline\hline       
 order & N    & RMS & $T_{\rm eff}$ & $log\,g$ & $v_{\rm rot}$\ sin$i$ \\
\hline  
 64    &  977 & 0.0396 & 1000 & 4.0 & 43 \\
 63    &  977 & 0.0437 & 1000 & 4.5 & 47 \\
 62    &  977 & 0.0386 &  800 & 5.0 & 41 \\
 61    &  976 & 0.0550 &  900 & 5.5 & 47 \\
 60    &  858 & 0.0533 &  900 & 3.5 & 37 \\
 59    &  977 & 0.0582 & 1100 & 5.0 & 40 \\
 58    &  977 & 0.0671 & 1300 & 5.0 & 35 \\
\hline                  
\end{tabular}
\end{table}

\begin{table}
\caption{SDSS J134646.45-003150.4. Synthetic and observed spectra comparison 
         for all orders: RMS, N, $T_{\rm eff}$, $log\,g$ and $v_{\rm rot}$\ sin$i$.}             
\label{table:5}      
\centering          
\begin{tabular}{lccccc}     % 41 columns 
\hline\hline       
 order & N    & RMS & $T_{\rm eff}$ & $log\,g$ & $v_{\rm rot}$\ sin$i$ \\
\hline  
 64    &  972 & 0.0544 & 1000 & 4.0 & 19 \\
 63    &  974 & 0.0554 & 1000 & 4.0 & 20 \\
 62    &  974 & 0.0500 &  900 & 4.5 & 17 \\
 61    &  974 & 0.0644 & 1000 & 5.0 & 25 \\
 60    &  862 & 0.0801 &  900 & 4.0 & 12 \\
 59    &  974 & 0.1185 &  900 & 4.0 & 13 \\
 58    &  973 & 0.1139 & 1200 & 4.5 & 13 \\
\hline                  
\end{tabular}
\end{table}

\begin{table}
\caption{2MASS J15530228+1532369. Synthetic and observed spectra comparison 
         for all orders: RMS, N, $T_{\rm eff}$, $log\,g$ and $v_{\rm rot}$\ sin$i$.}             
\label{table:6}      
\centering          
\begin{tabular}{lccccc}     % 41 columns 
\hline\hline       
 order & N    & RMS & $T_{\rm eff}$ & $log\,g$ & $v_{\rm rot}$\ sin$i$ \\
\hline  
 63    &  976 & 0.0507 & 1000 & 4.5 & 37 \\
 62    &  976 & 0.0582 &  800 & 5.0 & 21 \\
 61    &  976 & 0.0659 &  900 & 5.5 & 39 \\
 60    &  857 & 0.0563 &  900 & 4.0 & 27 \\
 59    &  976 & 0.0733 &  900 & 4.0 & 30 \\
 58    &  979 & 0.0863 & 1100 & 5.0 & 23 \\
\hline                  
\end{tabular}
\end{table}

\begin{table}
\caption{2MASS J12171110-0311131. Synthetic and observed spectra comparison 
         for all orders: RMS, N, $T_{\rm eff}$, $log\,g$ and $v_{\rm rot}$\ sin$i$}             
\label{table:7}      
\centering          
\begin{tabular}{lccccc}     % 41 columns 
\hline\hline       
 order & N    & RMS    & $T_{\rm eff}$ & $log\,g$ & $v_{\rm rot}$\ sin$i$ \\
\hline  
 63    &  843 & 0.0467 &  800 & 5.5 & 21 \\
 62    &  972 & 0.0581 &  900 & 4.0 & 32 \\
 61    &  972 & 0.0947 &  900 & 5.5 & 41 \\
 60    &  882 & 0.1777 &  900 & 4.0 & 21 \\
 59    &  972 & 0.0894 & 1000 & 4.5 & 27 \\
 58    &  972 & 0.0918 & 1100 & 4.5 & 29 \\
\hline                  
\end{tabular}
\end{table}

\begin{table}
\caption{GL 570D. Synthetic and observed spectra comparison 
         for all orders: RMS, N, $T_{\rm eff}$, $log\,g$ and $v_{\rm rot}$\ sin$i$.}             
\label{table:8}      
\centering          
\begin{tabular}{lccccc}     % 41 columns 
\hline\hline       
 order & N & RMS & $T_{\rm eff}$ & $log\,g$ & $v_{\rm rot}$\ sin$i$ \\
\hline  
 64    &  974 & 0.0327 &  900 & 4.5 & 27 \\
 63    &  974 & 0.0411 & 1000 & 4.5 & 34 \\
 62    &  974 & 0.0428 & 1000 & 4.0 & 41 \\
 61    &  974 & 0.0649 &  900 & 5.5 & 41 \\
 60    &  861 & 0.0675 &  900 & 4.0 & 26 \\
 59    &  974 & 0.0919 & 1000 & 4.5 & 26 \\
 58    &  974 & 0.0986 & 1000 & 5.0 & 21 \\
\hline                  
\end{tabular}
\end{table}

\begin{table}
\caption{2MASS J04151954-0935066. Synthetic and observed spectra comparison 
         for all orders: RMS, N, $T_{\rm eff}$, $log\,g$ and $v_{\rm rot}$\ sin$i$.}             
\label{table:9}      
\centering          
\begin{tabular}{lccccc}     % 41 columns 
\hline\hline       
 order & N    & RMS & $T_{\rm eff}$ & $log\,g$  & $v_{\rm rot}$\ sin$i$ \\
\hline  
 64    &  926 & 0.2029 &  800 & 3.5 & 43 \\
 63    &  981 & 0.1791 & 1000 & 3.5 & 33 \\
 62    &  982 & 0.1119 & 1000 & 4.0 & 44 \\
 61    &  987 & 0.1206 &  900 & 5.0 & 43 \\
 60    &  899 & 0.1539 & 1000 & 3.5 & 38 \\
 59    &  987 & 0.1081 &  900 & 4.5 & 29 \\
 58    &  984 & 0.1330 & 1000 & 5.0 & 25 \\
\hline                  
\end{tabular}
\end{table}

\section{Analysis}
\label{analysis}

In order to constrain the number of possible solutions provided by our large 
set of models, the root-mean-square $RMS(v_{\rm rot}{\rm sin}i, T_{\rm eff}, log\,g)$ is obtained for each 
model:

   \begin{equation}
     RMS(v_{\rm rot}{\rm sin}i, T_{\rm eff}, log\,g) = \sqrt{\frac{1}{N}\sum_{i=1}^{n}\left(F_\nu(i)-G_\nu(i)\right)^2\,},
   \end{equation}

where $i$ stands for i-pixel in the spectral axis, $F_\nu(i)$ and $G_\nu(i)$ 
correspond to the observed and modelled fluxes, respectively. 
We compute values of $RMS$ for the whole set of models and for 
all echelle orders independently, since different orders show 
different signal-to-noise ratios.

Tables \ref{table:1}-\ref{table:9} show the values of $T_{\rm eff}$,
log\,$g$, and $v_{\rm rot}$ corresponding to the model that minimizes 
the $RMS$ for each echelle order. 
The total number of pixels $N$ used in the comparison is also listed. 
For order 60, the wavelength range between 1.268 and 1.270 $\mu$m is excluded 
from the comparison because of the presence of strong telluric lines
that practically blocked any signal from the targets. To avoid border
effects, a few pixels at both sides of each echelle order have been
rejected.

\subsection{Average physical parameters}
\label{averageparam}

\begin{table}
\caption{Weights of echelle orders.}             
\label{table:10}      
\centering          
\begin{tabular}{lcc}     % 3 columns 
\hline\hline       
Order & Weight  & Remark \\
\hline    
66 &  0.3 & for all objects \\
65 &  0.4 & for all objects \\
64 &  0.6 & for all objects \\
63 &  1.0 & for all objects \\
62 &  1.0 & for all objects \\
61 &  1.0 & for all objects \\
60 &  0.2 & for 2MASS J12171110-0311131 \\
   &  0.6 & for 2MASS J04151954-0935066 \\
   &  1.0 & for the rest \\
59 &  1.3 & for all objects \\
58 &  1.1 & for all objects \\
57 &  0.8 & for all objects \\
\hline                  
\end{tabular}
\end{table}

We compute the {\em average} values of $T_{\rm eff}$, log\,$g$, and $v_{\rm rot}$ 
for each object in our sample by combining the various values obtained for the 
different echelle orders according to the following equations:

   \begin{equation}
     <T_{\rm eff}> = \frac{\sum_{j}\left(T_{eff,j}\frac{W_j}{\sigma_j}\right)}{\sum_{j}\frac{W_j}{\sigma_j}}\,,
   \end{equation}

   \begin{equation}
     <log\,g> = \frac{\sum_{j}\left(log\,g_j\frac{W_j}{\sigma_j}\right)}{\sum_{j}\frac{W_j}{\sigma_j}}\,,
   \end{equation}

   \begin{equation}
     <v_{\rm rot}{\rm sin}i> = \frac{\sum_{j}\left(v_{\rm rot}{\rm sin}i_j\frac{W_j}{\sigma_j}\right)}{\sum_{j}\frac{W_j}{\sigma_j}}\,,
   \end{equation}

where $\sigma_j$ refers to the minimum $RMS$ for the echelle
order $j$ (as listed in Tables \ref{table:1}-\ref{table:9}), 
and $W_j$ corresponds to the weight that we have assigned to 
the echelle order $j$ based on the signal-to-noise ratio of the 
observations. Table \ref{table:10} shows the various $W_j$, 
which are relative to order 61. Regarding the echelle order 60, 
we have given lower weights to J1217$-$0311 and J0415$-$0935 
because the correction for Earth telluric lines were not optimal.
Finally, the average values of $T_{\rm eff}$, log\,$g$, and 
$v_{\rm rot}$ are provided in Table \ref{table:parameters}, 
where we also include the spectral types of the targets 
following the unified scheme of Burgasser et al. (2006). 
Errors in $<T_{\rm eff}>$, $<log\,g>$ and $<v_{\rm rot}{\rm sin}i>$ 
were obtained from the standard deviation of the corresponding
 values derived for the various echelle orders.
We note that if no weights are taken into account, the solutions 
for $<T_{\rm eff}>$, $<log\,g>$ and $<v_{\rm rot}{\rm sin}i>$ do not 
change significantly, with differences with respect of those 
obtained using the weights $W_j$ of $\pm$40\,K (for T dwarfs later 
than T5 the difference is only $\pm$7 K), $\pm$0.1 dex, and 
$\pm$1 Km s$^{-1}$, which are much smaller than the quoted uncertainties.

\begin{figure}
   \centering
   \includegraphics[width=9.cm]{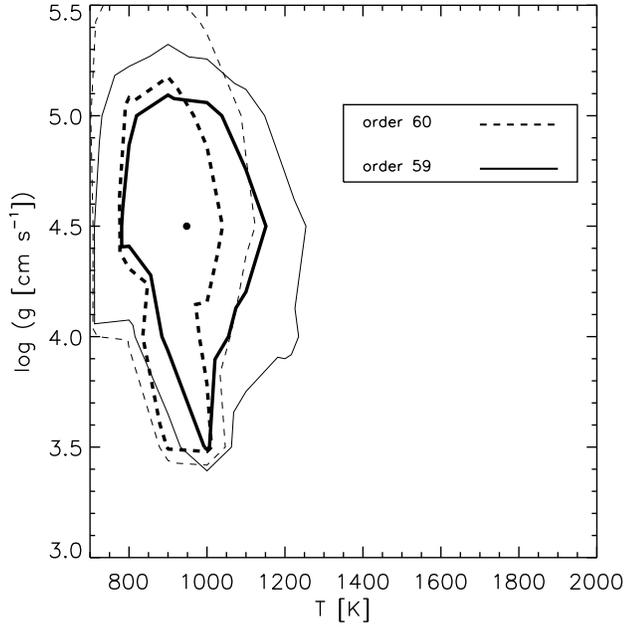}
\caption{Contours with RMS=1.25 (thick lines) and 1.5 (thin lines) times the value of the minimum RMS for orders 59 (solid lines) and 60 (dashed lines). Location of ($<T_{\rm eff}>$, $log\,g$) for the minimum RMS is also indicated (dot).}
              \label{FigCdB5}%
   \end{figure}

Fig. \ref{FigCdB5} shows, as an example, some contours of equal RMS (Root mean square) around the solution ($<T_{\rm eff}>$, $log\,g$) for GL570D, with $v_{\rm rot}$=32 Km s$^{-1}$. The isocontour of the RMS of 1.25 times
the minimum RMS enclose values of $log\,g$ of $\approx\pm$0.7 dex and $T_{\rm eff}$ of $\approx\pm$150\,K around the solution. In general, the isocontours of the RMS enclose a unique solution with a broad range of values of gravity and temperature around the solution.

\subsection{Best modelled spectra}
\label{bestmodels}

We have computed the modelled spectra determined from 
the average effective physical parameters determined in Sect. \ref{averageparam}
(hereinafter, {\it best} modelled spectra). 
We have used a bilinear interpolation to compute the synthetic models corresponding 
to $<T_{\rm eff}>$ and $<log\,g>$ from our {\em AMES-COND} model grid. This
bilinear interpolation has been done using the logarithm of the synthetic fluxes. 
We first proved that a linear interpolation in both axes was appropriate. 
We fixed $T_{\rm eff}$ at 1000\,K and used the corresponding 
synthetic models at log\,$g$=4.0 and log\,$g$=5.0
to compute interpolated spectra at log\,$g$=4.5. We also fixed log\,$g$=4.5 and
used the corresponding synthetic models at $T_{\rm eff}=$900\,K and 1100\,K
to compute the interpolated spectra at $T_{\rm eff}=$1000\,K. We then applied the 
same steps described in Sect 2.3 to compare the synthetic spectra to our observations.
The differences between the interpolated and synthetic models are
typically of a few per cent. The largest differences of 10\% observed between the spectra 
were obtained for the echelle order 57 with a null rotational broadening.
For the rest of the echelle orders the differences were generally below 1\%.

\begin{figure}
   \centering
   \includegraphics[width=9.cm]{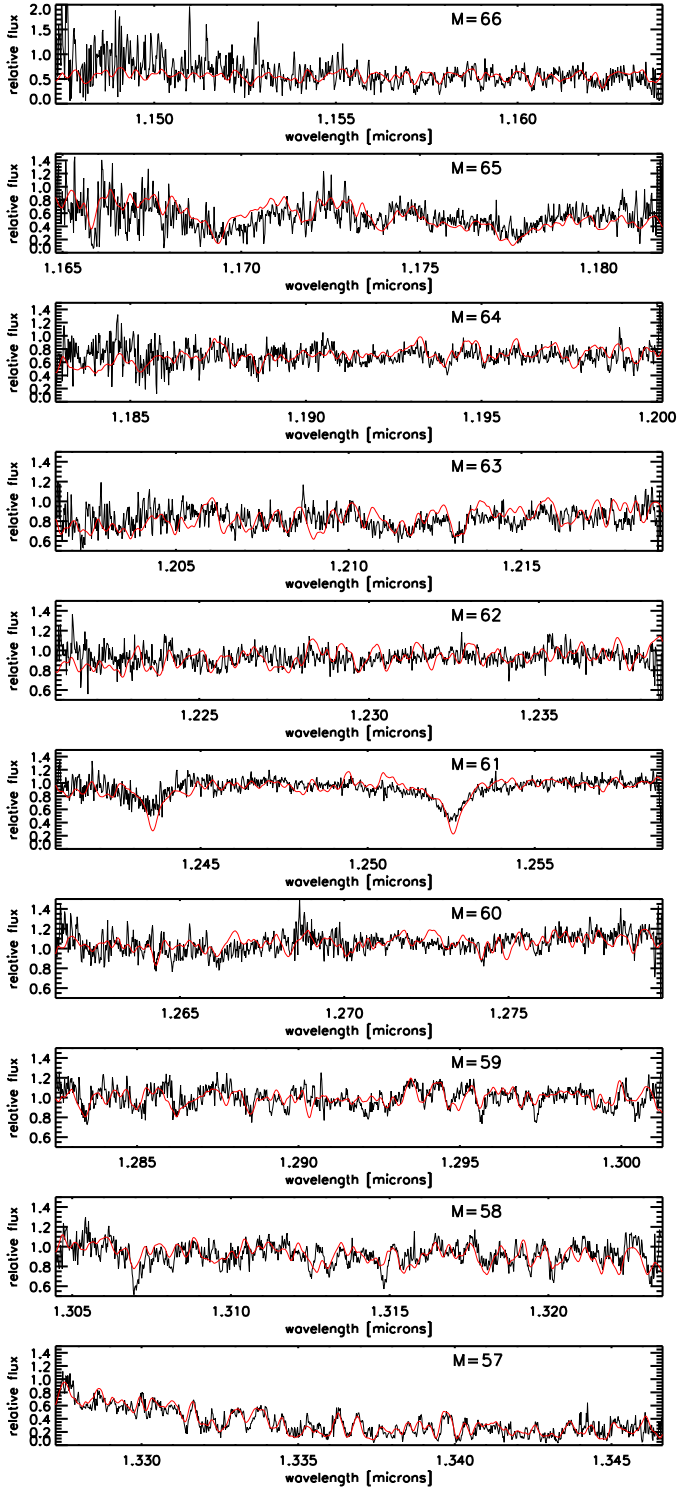}
\caption{SDSSp J125453.90-012247.4. Black and red lines correspond to the observed 
         and modelled spectra, respectively. The model is for an effective temperature
         of 2000 \,K and a surface gravity of 10$^{4.3}$ cm s$^{-2}$.}
              \label{FigCdB6}%
   \end{figure}

\begin{figure}
   \centering
   \includegraphics[width=9.cm]{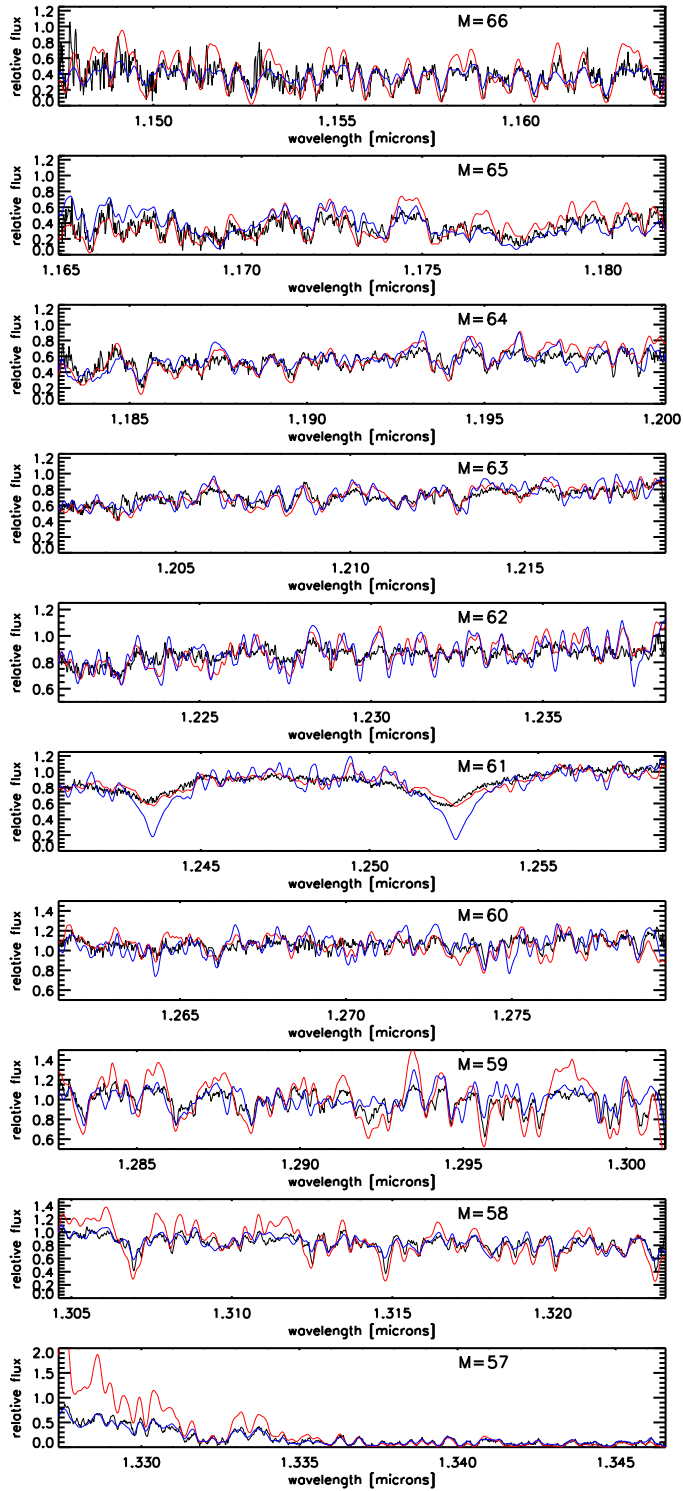}
\caption{2MASS J05591914-1404488. Black line corresponds to the observed 
         spectrum, and red and blue lines correspond to modelled 
         spectra with the same $g$=10$^{4.9}$ cm s$^{-2}$ and 
         $T_{\rm eff}$=1002, and 1700 \,K.}
              \label{FigCdB7}%
   \end{figure}

\begin{figure}
   \centering
   \includegraphics[width=9.cm]{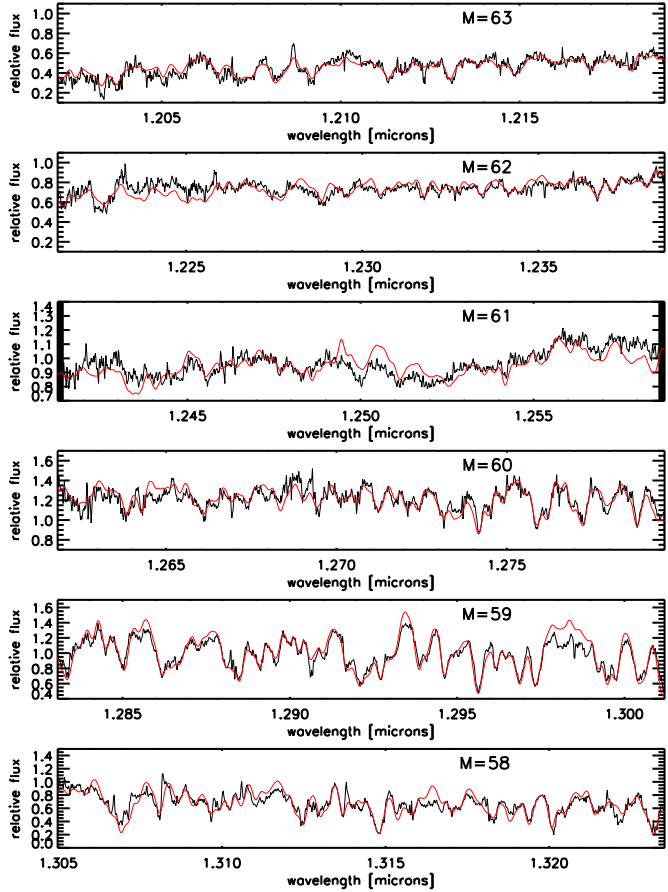}
\caption{2MASS J15530228+1532369. Black and red lines correspond to the observed 
         and modelled spectra, respectively. The model is for an effective temperature
         of 941 \,K and a surface gravity of 10$^{4.6}$ cm s$^{-2}$.}
              \label{FigCdB11}%
   \end{figure}

\begin{figure}
   \centering
   \includegraphics[width=9.cm]{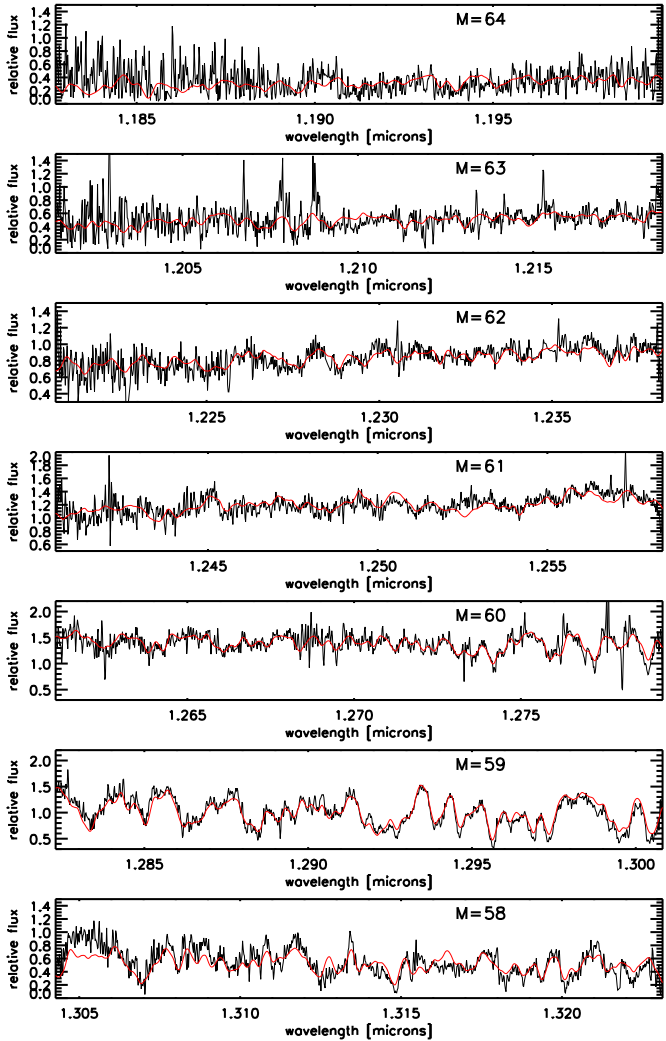}
\caption{2MASS J04151954-0935066. Black and red lines correspond to the observed 
         and modelled spectra, respectively. The model is for an effective temperature
         of 947 \,K and a surface gravity of 10$^{4.3}$ cm s$^{-2}$.}
              \label{FigCdB14}%
   \end{figure}

Figs. \ref{FigCdB6}-\ref{FigCdB14} 
show the observations and modelled spectra for our list of T dwarfs.

\section{Results}
\label{results}

\begin{table*}
\caption{Projected rotational velocity, temperature, surface gravity, mass and age of T dwarfs.}             
\label{table:parameters}      
\centering          
\begin{tabular}{l c c c c c c c c c c}     % 5 columns 
\hline\hline       
Object                    & Spectral Type & $v_{\rm rot} sin i$ & $<T_{\rm eff}>$ & PREV. & REF. & $<log\,g>$ & PREV. & REF.  & $M$ & $Age$ \\
         &          & {\em Km s$^{-1}$}   &   {\em K} &   {\em K} & & {\em (cm s$^{-2}$)} & {\em (cm s$^{-2}$)} & &  $M_{J}$ & {\em Gyr} \\
\hline    
SDSSp J125453.90-012247.4 & T2   & 34$\pm$7  & 2007$\pm$662 & 1200                & 1 & 4.3$\pm$1.0 & 5.0          & 1 & - & - \\
2MASS J05591914-1404488   & T4.5 & 28$\pm$5  & 1002$\pm$278 & 1200                & 1 & 4.9$\pm$0.8 & 5.5          & 1 & $30^{+45}_{-23}$ & $1.0^{+9}_{-0.9}$  \\
2MASS J15031961+2525196   & T5   & 36$\pm$5  & 1009$\pm$217 & 1178                & 2 & 4.6$\pm$0.8 & $\approx$5.0 & 5 & $18^{+37}_{-13}$ & $0.4^{+4.6}_{-0.3}$  \\
SDSS J162414.37+002915.6  & T6   & 42$\pm$5  &  980$\pm$163 & 1002$^{+98}_{86}$   & 3 & 4.8$\pm$0.7 & $\approx$5.0 & 5 & $23^{+52}_{-16}$ & $0.6^{+9.4}_{-0.5}$  \\
SDSS J134646.45-003150.4  & T6.5 & 16$\pm$5  &  990$\pm$107 & 960-1020            & 4 & 4.1$\pm$0.4 & 5.0-5.2      & 4 & $10^{+10}_{-5}$   & $0.1^{+0.4}_{-0.08}$ \\
2MASS J15530228+1532369   & T7   & 30$\pm$7  &  941$\pm$138 & 893                 & 2 & 4.6$\pm$0.6 & $\approx$5.0 & 5 & $18^{+30}_{-11}$  & $0.4^{+3.6}_{-0.3}$  \\
2MASS J12171110-0311131   & T7.5 & 29$\pm$7  &  922$\pm$103 & 860-880             & 4 & 4.8$\pm$0.7 & 4.7-4.9      & 4 & $23^{+52}_{-16}$ & $0.8^{+9}_{-0.7}$  \\
GL 570D 		  & T7.5 & 32$\pm$8  &  948$\pm$53  & 780-820             & 4 & 4.5$\pm$0.5 & 5.1          & 4 & $15^{+15}_{-9}$  & $0.3^{+1.7}_{-0.2}$  \\
2MASS J04151954-0935066   & T8   & 36$\pm$7  &  947$\pm$79  & 740-760             & 4 & 4.3$\pm$0.7 & 4.9-5.0      & 4 & $10^{+20}_{-6}$  & $0.1^{+1.9}_{-0.09}$ \\
\hline                  
\end{tabular}
\begin{list}{}{}
\item[${\mathrm{1}}$] Cushing et al. (2008) \item[${\mathrm{2}}$] Estimated $T_{\rm eff}$ derived from the $T_{\rm eff}$--spectral type relation in Looper et al. (2008) \item[${\mathrm{3}}$] Vrba et al. (2004) \item[${\mathrm{4}}$] Burgasser et al. (2006) \item[${\mathrm{5}}$] Knapp et al. (2004)
\end{list}
\end{table*}

\subsection{Near-infrared absorption features of T dwarfs}

As observed by McLean et al. (2007), the $J$-band spectral morphology 
at $R\sim20,000$ of T dwarfs shows a dense population of weak absorption 
features and a few relatively strong lines. The fine-scale spectral 
structure is mainly associated with H$_2$O, and the strong 
atomic features are doublets of K\,{\sc i} that appear in orders 61 and 
65. The strong K\,{\sc i} doublet in order 61, at wavelengths 1.2436 and 
1.2525 $\mu$m, deepen towards earlier Ts (see Fig. \ref{FigCdB1}). 
We note that the lines of this doublet in SDSS J134646.45-003150.4 
are stronger than what would be expected according to 
its spectral type (T6.5).

FeH lines in orders 62 and 63 are weaker towards later Ts. Some
absorption around 1.222 $\mu$m may be present even at T4.5.
Those at 1.24637 and 1.24825 $\mu$m FeH in order 61 are 
not visible in T dwarfs. Order 64 is dominated by sharp and deep
H$_2$O absorption features. This is also the case of order 66,
and more dramatically for order 57, with the strongest H$_2$O 
absorption in the observed range. 
Order 65 shows the other K\,{\sc i} doublet and changes
with spectral type. It also contains many strong intrinsic transitions 
of hot H$_2$O, like the feature at 1.175 $\mu$m.

\begin{figure}
   \centering
   \includegraphics[width=9.cm]{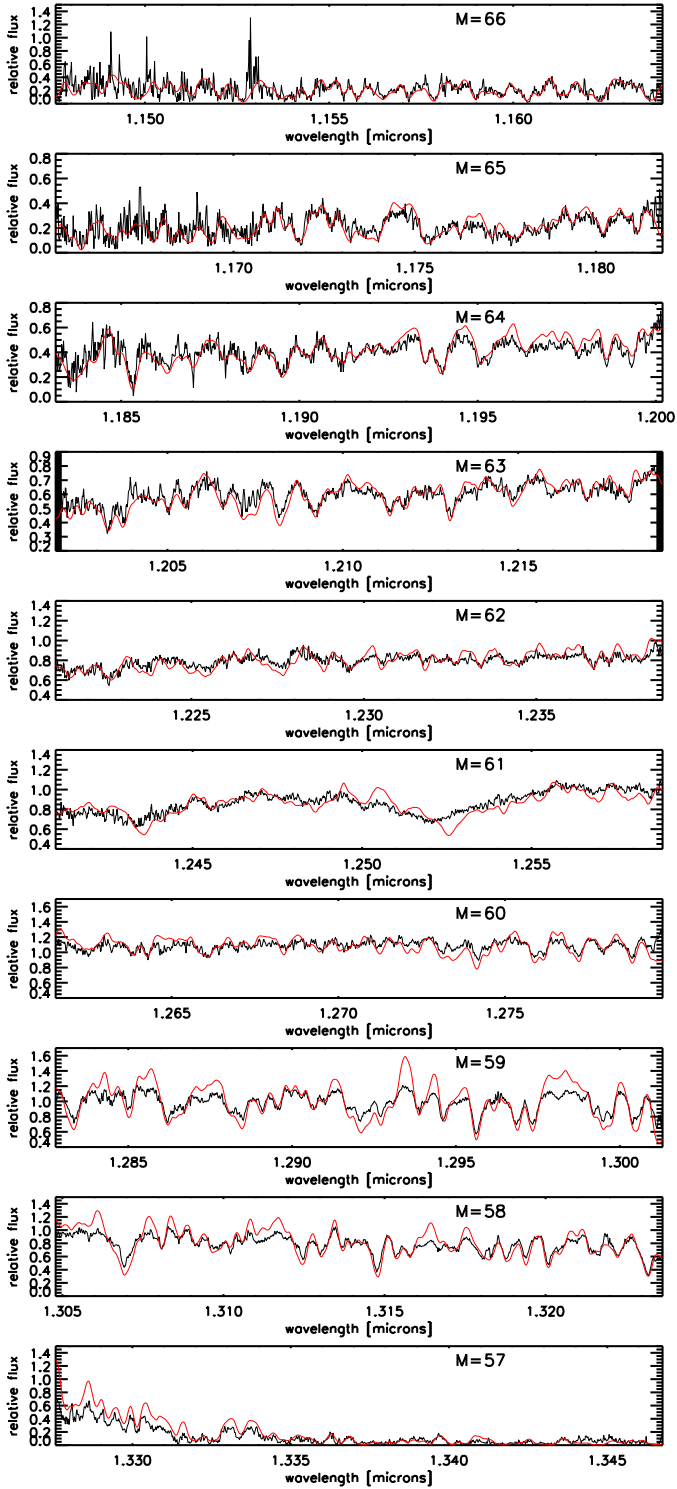}
\caption{2MASS J15031961+2525196. Black and red lines correspond to the observed and
         modelled spectra, respectively. The model is for an effective temperature
         $T_{\rm eff}$=1009 \,K and $g$=10$^{4.6}$ cm s$^{-2}$.}
              \label{FigCdB8}%
   \end{figure}

\begin{figure}
   \centering
   \includegraphics[width=9.cm]{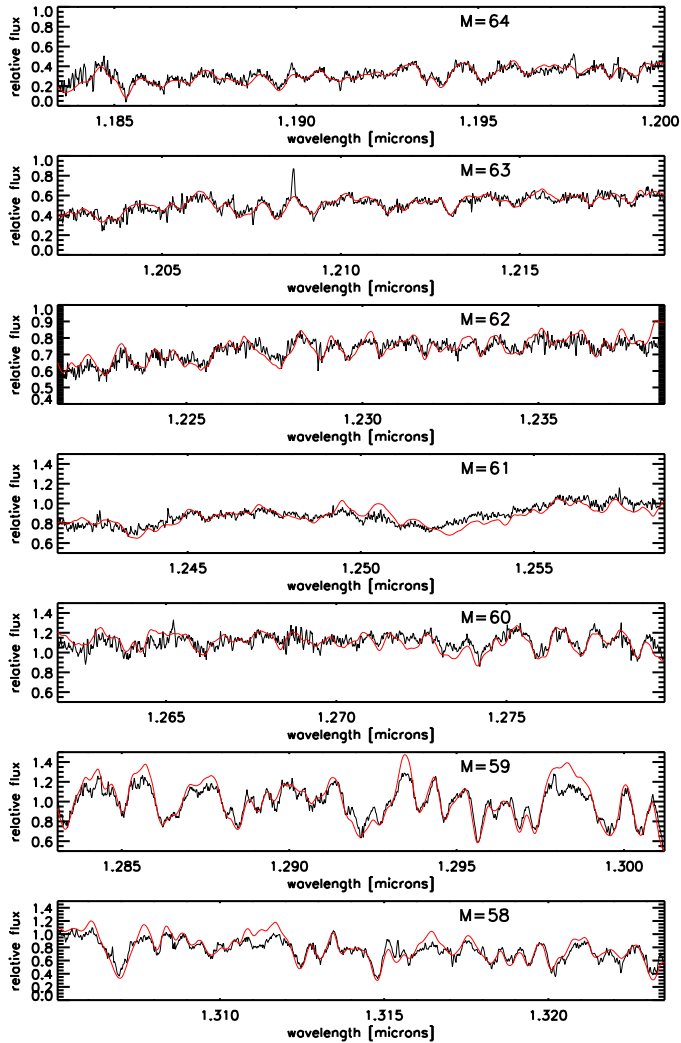}
\caption{SDSS J162414.37+002915.6. Black and red lines correspond to the observed 
         and modelled spectra, respectively. The model is for an effective temperature
         of 980 \,K and a surface gravity of 10$^{4.8}$ cm s$^{-2}$.}
              \label{FigCdB9}%
   \end{figure}

\begin{figure}
   \centering
   \includegraphics[width=9.cm]{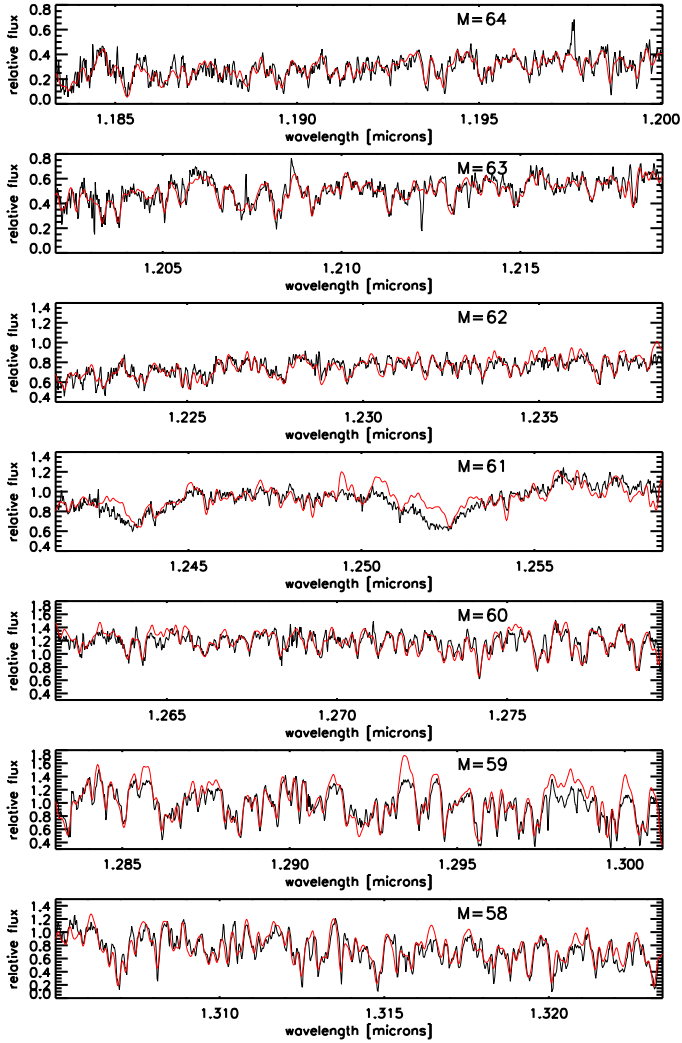}
\caption{SDSS J134646.45-003150.4. Black and red lines correspond to the observed 
         and modelled spectra, respectively. The model is for an effective temperature
         of 990 \,K and a surface gravity of 10$^{4.1}$ cm s$^{-2}$.}
              \label{FigCdB10}%
   \end{figure}

\begin{figure}
   \centering
   \includegraphics[width=9.cm]{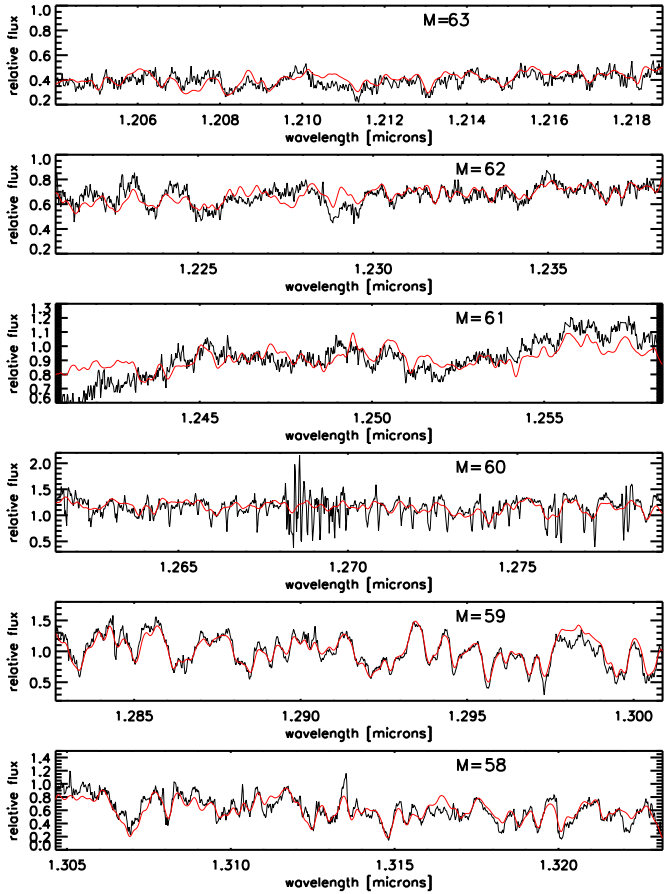}
\caption{2MASS J12171110-0311131. Black and red lines correspond to the observed 
         and modelled spectra, respectively. The model is for an effective temperature
         of 922 \,K and a surface gravity of 10$^{4.8}$ cm s$^{-2}$.}
              \label{FigCdB12}%
   \end{figure}

\begin{figure}
   \centering
   \includegraphics[width=9.cm]{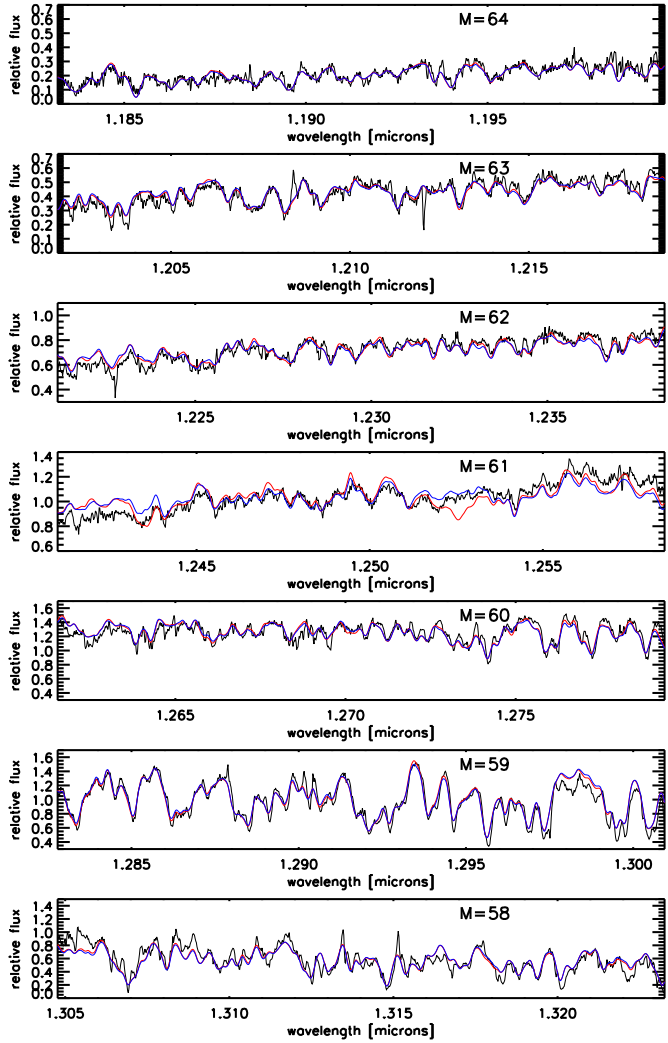}
\caption{GL 570 D. Black line corresponds to the observed spectrum,
         and red and blue lines correspond to modelled spectra with the same
         $g$=10$^{4.5}$ cm s$^{-2}$ and $T_{\rm eff}$=948 \,K and 900 \,K.}
              \label{FigCdB13}%
   \end{figure}

In general, our observed spectra are well reproduced by the 
best modelled spectra. In a few cases we have found some small 
differences. This is the case of 2MASS J15031961+2525196 (T5) (see
Fig. \ref{FigCdB8}), where the best model is found for $log\,g$=4.6 
and $T_{\rm eff}$=1009\,K. The observed K\,{\sc i} doublet in order 61 is 
not perfectly reproduced by the model. We also note some
differences in orders 59 and 58. Small differences in these orders 
are also present in SDSS J162414.37+002915.6 (T6) (see Fig. 
\ref{FigCdB9}). These are much smaller in later spectral types 
(e.g., SDSS J134646.45-003150.4 (T6.5), see Fig. \ref{FigCdB10}).
For late T dwarfs, with faint K\,{\sc i} doublets, the models provide
an excellent match to the observations for all the orders, with a
few exceptions such as the order 61 in 2MASS J12171110-0311131
(see Fig. \ref{FigCdB12}), where the model is flat towards the 
blue region. We note that the observed spectrum may be affected 
by errors in the flat-fielding. Note also that there is 
a faint line of K\,{\sc i} at 1.2525 $\mu$m in the best modelled spectra 
for order 61 of GL 570 D that disappears when considering an 
effective temperature 48 \,K lower (see Fig. \ref{FigCdB13}).

The two earlier T dwarfs of our sample, SDSSp J125453.90-012247.4 
and 2MASS J05591914-1404488, present significant differences 
between the best modelled and observed spectra. 
The modelled spectrum corresponding to the {\it average} values 
($T_{\rm eff}$=1002\,K and $log\,g$=4.9) in 2MASS J05591914-1404488
fails to mimic order 57 with many water vapor lines. The
K\,{\sc i} doublet at 1.2436 and 1.2525 $\mu$m, very sensitive
to temperature, is also difficult to reproduce. Order 57 is clearly 
better modelled with a higher temperature $T_{\rm eff}$=1700\,K.

\subsection{Physical parameters of T dwarfs}

\subsubsection{Rotation, effective temperature and surface gravity}

For our list of nine T dwarfs, we have determined their projected
rotational velocity ($v_{\rm rot} sin i$), effective temperature and 
surface gravity from the comparison of the high resolution spectra 
with the {\em AMES-COND} models (see Table \ref{table:parameters}).
The rotational velocities determined here from modelled spectra are 
in good agreement (i.e., within 1-$\sigma$ uncertainties) 
with those obtained by ZO06, who used J134646.45-003150.4 as a non-rotator
reference to measure the $v_{\rm rot}{\rm sin}i$ parameter of the
sample (i.e., they assumed $v_{\rm rot}{\rm sin}i$=0 for this object). 
With the exception of 2MASS 12171110-0311131, our values are, indeed, 
systematically higher by a few Km s$^{-1}$, which is expected since 
the template used by ZO06 has a small rotation of 16\,Km\,s$^{-1}$
according to our analysis.

The effective temperatures of the T dwarfs with spectral types later
than T5 are between 922 and 1009 \,K, with errors between $\approx$50 and 200\,K
(see Table \ref{table:parameters}). 
The surface gravities of those objects lie between $10^{4.1}$ 
and $10^{4.9}\,cm s^{-2}$, and errors are of $\approx$0.7 dex. 
In Sections \ref{dissefftemp} and \ref{disssurfgrav} we
discuss our results on $T_{\rm eff}$ and $log\,g$.

\subsubsection{Mass and age}

\begin{figure}
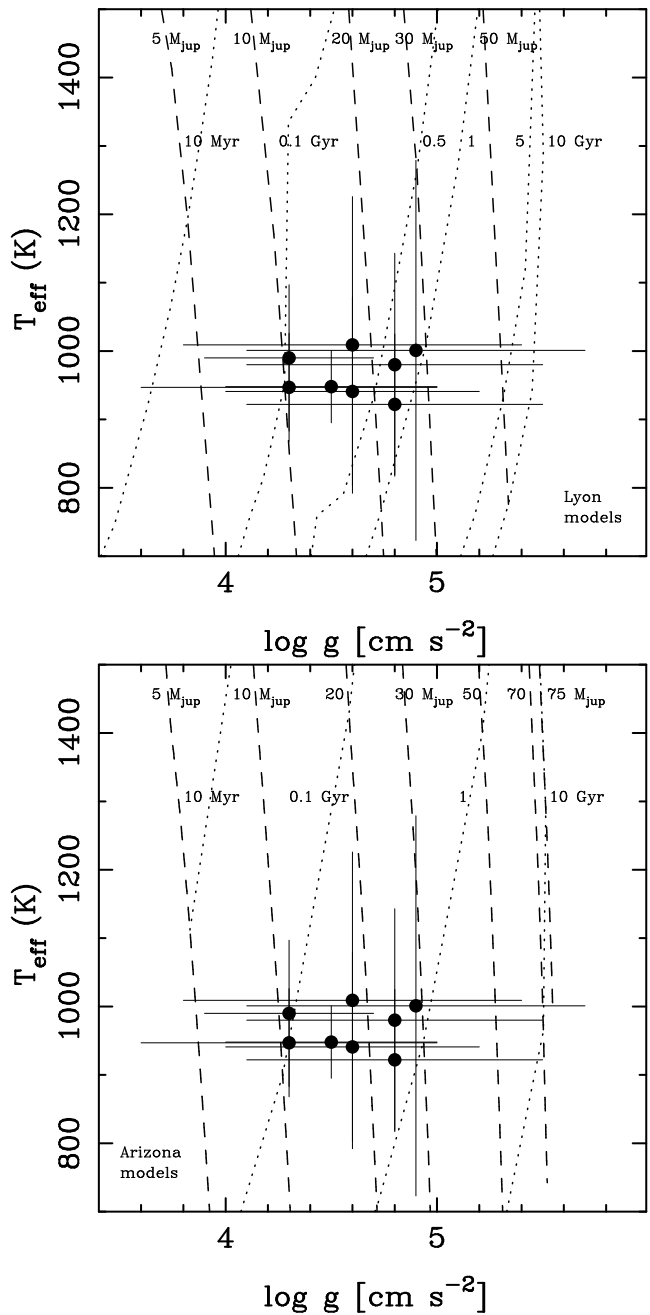

   \centering
   \includegraphics[width=8.5cm]{FigCdB15a.ps}
   \includegraphics[width=8.5cm]{FigCdB15b.ps}
\caption{Diagram $T_{\rm eff}$ versus $log\,g$ with the values we have obtained
         for our sample of T dwarfs (SDSSp J125453.90-012247.4 is not included) 
         and the solar metallicity models by Lyon ({\em top}) and Arizona ({\em bottom}). 
         The models provide isochrones from 10 Myr to 10 Gyr (dashed lines) and 
         curves of constant mass (dotted lines).}
              \label{FigCdB15}%
   \end{figure}

We have estimated the mass and age of our T dwarfs using state-of-the-art 
evolutionary models.
Our spectroscopically derived values $<T_{\rm eff}>$ and $<log\,g>$ are plotted 
in the two panels of Fig.~\ref{FigCdB15}. The T2V dwarf J1254$-$0122 is 
discarded from the Figure since the modelled spectra found here do not 
reproduce the observations satisfactorily.
Therefore, only mid- and late-T dwarfs are shown in 
the $T_{\rm eff}$ versus $log\,g$ diagram, which we remark is independent 
of the distance to the sources. Fig.~\ref{FigCdB15} ({\em left}) shows,
overplotted on the data, the solar metallicity models by 
the Lyon group (Baraffe et al. 2003), with isochrones (from 10\,Myr to 10\,Gyr) 
and curves of constant mass (5, 10, 20, 30, and 50\,$M_{\rm J}$). 
Fig.~\ref{FigCdB15} ({\em right}) shows our data with 
the solar metallicity models by the Arizona group (Burrows et al. 1997): 
isochrones (10 Myr, 0.1, 1 and 10\,Gyr) and curves of constant mass (5, 10, 20, 
30, 50, 70 and 75 \,$M_{\rm J}$).

Table \ref{table:parameters} shows the values 
of mass and age for our sample of T dwarfs that are derived from the 
$T_{\rm eff}$ versus $log\,g$ diagram using the models of the Lyon group. 
The mass and age uncertainty intervals are inferred from the error bars in 
the effective temperature and surface gravity.
Summarizing, all these objects have a mass in the interval 
5--75 \,$M_{\rm J}$, thus confirming their very likely substellarity, 
and likely ages that interestingly seem to be younger than 
the solar system. We note, however, that considering the large
error bars, most of the T dwarfs in our sample may have an age consistent
with a few to several Gyr. For GL\,570\,D and 2MASS J04151954-0935066 
we have derived an age upper limit of 1--2 Gyr, and only for 
SDSS J134646.45-003150.4 the upper limit is below 1 Gyr.

\section{Discussion}
\label{discussion}

\subsection{Modelled and observed spectra}
\label{dissfeatures}

Most of the lines in a T dwarf are due to water vapor. A few atomic 
lines can be identified, but some of them, e.g. the Rb I line, are 
very weak compared to the haze of molecular lines.

In general, the modelled spectra match remarkably well the 
high-resolution observed spectra. We note only that the profiles 
of the strong K \,{\sc i} lines are not so well matched by the models. 
This is likely due to the approximations used for the intrinsic 
line profiles in the models, most importantly estimates of the 
damping constants. For example, Allard et al. (2003) and Johnas 
et al. (2007) have demonstrated the importance of detailed line
profiles (not just damping constants) for optical alkali resonance 
lines in cool sub-stellar objects, however, such profiles are 
not yet available for the IR alkali lines. Therefore, the profiles 
of these atomic lines cannot be modelled accurately, and so cannot 
be used to obtain reliable parameter estimates.
From qualitative considerations, however, the models predict that 
for surface temperatures of 900-1000K, the atomic lines of K\,{\sc i} 
strengthen with decreasing gravity, as was pointed out by 
Knapp et al. (2004). SDSS J134646.45-003150.4 shows the K\,{\sc i} 
doublet at 1.25 $\mu$m, in contrast to SDSS J162414.37+002915.6 
and 2MASS J15530228+1532369, 
which have  similar spectral types. From our analysis, we have found 
the smallest $log\,g$ for SDSS J134646.45-003150.4 in our sample. This
is also the object with the lowest estimated age. 

The FeH lines are difficult to model, in part due to their relatively 
poorly known oscillator strength and possible errors in the chemistry. 
The fact that they become weaker with lower effective
temperatures (like those of our sample) is due to the condensation of iron, 
which reduces the number density of FeH, resulting in an 
important test of the treatment of condensation in the equation of state.

CH$_4$ bands are expected and have been observed in T dwarfs 
(Burgasser et al. 2002) but, in the spectral range considered 
here, are much weaker than the water vapor lines and so cannot 
be identified with confidence.

At higher effective temperatures, as for SDSSp J125453.90-012247.4 (T2) 
and 2MASS J05591914-1404488 (T4.5), the models appear to be generally 
less consistent with the observations. This is likely due to the presence 
of remnants of dust clouds, floating around in upper layers, that are
not included in the modelled spectra (see Ruiz et al. 1997, Ackerman \& 
Marley 2001, Burgasser et al. 2002, Burrows et al. 2006, 
Cooper et al. 2003, Helling et al. 2008). Cushing et al. (2008) have
computed the properties of SDSSp J125453.90-012247.4 and 
2MASS J05591914-1404488 from the comparison of low and intermediate 
resolution spectra in the 0.95-14.5 $\mu$m wavelength range and 
synthetic spectra.
They found that SDSSp J125453.90-012247.4 has condensate clouds that
are thicker than those in 2MASS J05591914-1404488, which may explain the
difference in the spectra of these overluminous T dwarfs. Cushing et al. 
also discussed the possible unresolved binarity in both objects. They
determined $T_{\rm eff}\sim$1200\,K ($<$1150\,K), $log\,g\sim$4.8 ($<$5.38)
and an age of $\sim$0.4 ($<$10) Gyr for 2MASS J05591914-1404488 assuming
it is a single object (equal mass binary). The same temperature of 
$\sim$1200\,K is found for SDSSp J125453.90-012247.4. Our values of
$T_{\rm eff}$, $log\,g$ and age of 2MASS J05591914-1404488 
(see Table \ref{table:parameters}) are in agreement with those 
of Cushing et al.

We will present a further study of the effects of the metallicity 
on 2MASS J05591914-1404488 using a new set of synthetic models 
(del Burgo et al., in preparation). The new models will include a 
completely new equation of state and (where possible and available) 
improved line data. Preliminary tests with the new equation 
of state show great improvements in the M dwarf regime. In addition, 
a physical model for the dust cloud formation coupled to the 
structure of the atmosphere is being developed; this is required 
at least for the transition T$\to$L$\to$M.

\subsection{Effective temperature}
\label{dissefftemp}

For the two earliest T dwarfs, 2MASS J05591914-1404488
and especially SDSSp J125453.90-012247.4, we found that the models 
used here cannot reproduce the observed spectra so well. 
The T4.5 dwarf 2MASS J05591914-1404488 has historically been 
considered enigmatic (see Vrba et al. 2004). Our best modelled 
spectra with $<T_{\rm eff}>$ of 1002$\pm$278\,K is, taking into
account the errors, consistent with the estimate of 1231 \,K from the 
relation of Looper et al. and the value of 1200\,K found by 
Cushing et al. (2008).
Our best modelled spectrum of SDSSp J125453.90-012247.4 indicates an 
effective temperature above 2000\,K, which is much higher than that 
found by Vrba et al. (2004) and Cushing et al. (2008), but 
consistent (i.e. within the errors) with a temperature around 1500 \,K.
This value is similar to the estimate of 1370 \,K in the relation of
Looper et al.

We also find some differences for the latest T dwarfs. Our values
of $<T_{\rm eff}>$ for GL 570 D and 2MASS J04151954-0935066 are
about 150\,K higher than those from the literature 
(see \ref{table:parameters}). 
Geballe et al. (2001) use R=400 spectra and accurate photometry
of GL 570D to determine that $T_{\rm eff}$ of 784-824 \,K and 
$log\,g$=5.00-5.27 ($cm s^{-2}$), assuming an age of 2-5 Gyr.

   \begin{figure}
   \centering
   \includegraphics[width=9.cm]{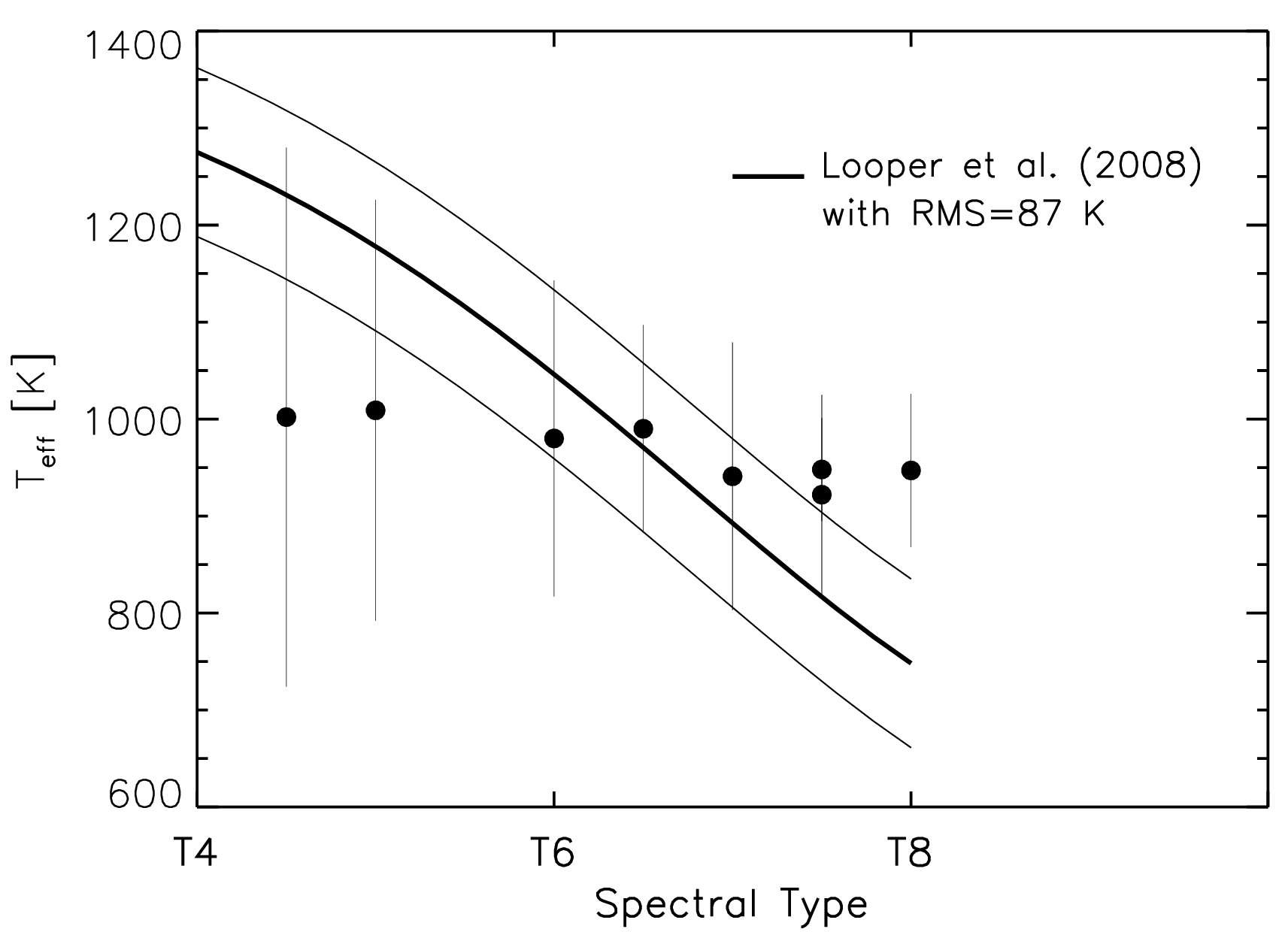}
\caption{$T_{\rm eff}$ versus spectral type for our T dwarfs (only SDSSp J125453.90-012247.4 is not included) and the $T_{\rm eff}$--spectral type relation of Looper et al. (2008).}
              \label{FigCdB16}%
   \end{figure}

Fig. \ref{FigCdB16} shows $T_{\rm eff}$ versus spectral 
type for the eight latest T dwarfs of our sample, and the 
$T_{\rm eff}$--spectral type relation found in Looper et al. (2008).
Our values appear to be rather flat from T4.5 down to T8, which
clearly contrasts with the trend deli\-neated by Looper et
al. (2008). We note, however, that such a trend is still compatible
with our measurements if error bars are taken into account. The
apparent constant temperature derived in our work may be due to a
degeneration in the method (synthetic and observed dataset) that, despite 
the high spectral resolution of the observations, is not accurately 
sensitive to $T_{\rm eff}$ and $log\,g$, partly due to the small wavelength 
range coverage of the data. Cushing et al. (2008) found that the values of
$T_{\rm eff}$ obtained from J-band low-resolution spectra are consistent
with those derived from fitting the full SED, using a model grid
with temperature steps of 100\,K.

\begin{figure}
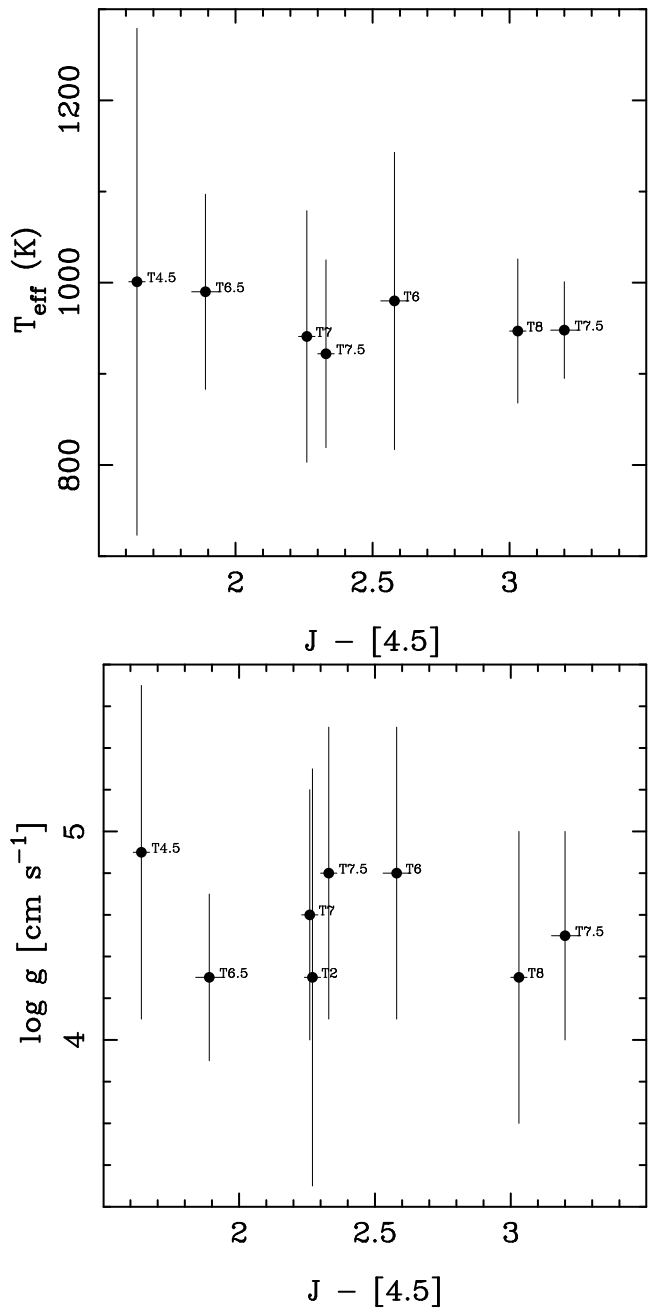

   \centering
   \includegraphics[width=8.5cm]{FigCdB17a.ps}
   \includegraphics[width=8.5cm]{FigCdB17b.ps}
\caption{$T_{\rm eff}$ versus colour J-[4.5] ({\em top}) and $log\,g$ versus J-[4.5] ({\em bottom}).}
              \label{FigCdB17}%
   \end{figure}

Fig. \ref{FigCdB17}({\it left}) shows
$<T_{\rm eff}>$ as a function of the colour $J-[4.5]$ for the T dwarfs
later than T4.5 (photometric data are compiled from Patten et al. 2006). 
These objects display a wide range of colours ($\sim$2 mag), 
however, our derived temperatures differ by less than 100\,K. It
seems difficult to reconcile such a large colour range with just one
value of temperature, unless other atmospheric parameters (like
metallicity, cloud coverage, and others) are taken into account.

\subsection{Surface gravity}
\label{disssurfgrav}

Our values of $log\,g$ are in good agreement with those of Knapp et al. (2004) 
(see Table \ref{table:parameters}), whose log$g$ values were derived 
from the comparison of observed H-K colors to model predicted H-K
colors. The largest difference is for the T8 dwarf 2MASS J04151954-0935066, 
where the estimate of Knapp et al. (2004) is 0.7 dex higher than our value.
Fig. \ref{FigCdB17}({\it right}) shows $log\,g$ versus 
infrared colour $J-[4.5]$, where no obvious trend is apparent.

\subsection{Mass and age}

The estimated masses in the T dwarfs of our sample are in good
agreement with the model fit values of Burgasser et al. (2004)
for three of the common objects (SDSS J162414.37+002915.6, 
2MASS J15530228+1532369, 2MASS J04151954-0935066). The only
exception is GL570 D. Burgasser et al. (2004) find a mass
of $<$0.001 $M_\odot$, i.e., $<$1 $M_J$, and also provide
a expected value of 30-50 $M_J$. We find a value of 15 $M_J$,
with an upper limit of 30 $M_J$ that is the same as the lower 
expected value given by Burgasser et al. Our lower limit (6 $M_J$)
is several times higher than the fitted value of Burgasser et al. (2004).

Our result on the {\em apparent young age} of the field T dwarfs (see 
Table \ref{table:parameters}) is in agreement with recent kinematical 
studies based on proper motions and space velocities by 
Bannister \& Jameson (2007) and Zapatero Osorio (2007). The latter 
authors found that about 40\%~of the L and T-type population of the 
solar neighborhood may have an age close to that of the Hyades cluster 
(around 600\,Myr), and that the brown dwarf population is kinematically 
younger than solar-type to early-M stars with likely ages in the 
interval 0.5--4\,Gyr. We also note that our upper limit of 2 Gyr for 
the widely studied object GL 570 D agrees with the lower limit assumed 
by Geballe et al. (2001).

\section{Conclusions}
\label{conclusions}

We conclude that the high resolution spectra corresponding to T dwarfs
with spectral types later than T5 are well reproduced by the 
{\em AMES-COND} solar metallicity models provided by the PHOENIX code.
The models reproduce in detail many faint absorption
features in the high resolution $J$-band spectra, which are mainly
due to water vapor. There are also strong K\,{\sc i} lines, which
turn out to be more difficult to model due to uncertain damping constants.
The temperature and surface gravity determined from the comparison of
the modelled and observed spectra are consistent with those found in the 
literature. We find a marked flat behaviour of $T_{\rm eff}$ with spectral 
type, although a possible gradient is compatible with the errors. 
High resolution spectroscopy ($R\sim20,000$) in the $J$-band and
{\em AMES-COND} models seems to be insufficient to show the existence of 
a possible gradient in effective temperature from early to late T dwarfs.
The comparison between the spectroscopically derived
$T_{eff}$ and $log g$ of our targets and the evolutionary models by the 
Lyon and Arizona groups yields ages in the range 0.1--5 Gyr and masses 
between $\sim$5 and 75 $M_{J}$ for the target sample with spectral 
types $\ge$T5. For the earlier type dwarfs the spectral models do 
not provide suitable fits, which is likely due to the presence of 
condensate clouds that are not incorporated in the models.

\begin{acknowledgements}
The authors would like to thank the referee for useful comments and
Carlos Allende for lively discussions.
The data used here were obtained at the W. M. Keck Observatory, which is
operated as a scientific partnership between the California Institute of
Technology, the University of California, and the National Aeronautics 
Space Administration.
Support for this project has been provided by the Spanish Ministry of
Science via project AYA2007-67458 and by a NASA-Keck analysis grant
provided by the Jet Propulsion Laboratory.
This work was also supported by the DFG via Graduiertenkolleg 1351.
Some of the calculations presented here were performed at the H\"ochstleistungs
Rechenzentrum Nord (HLRN); at the NASA's Advanced Supercomputing Division's
Project Columbia, at the Hamburger Sternwarte Apple G5 and Delta Opteron 
clusters financially supported by the DFG and the State of Hamburg; and at the
National Energy Research Supercomputer Center (NERSC), which is supported by
the Office of Science of the U.S. Department of Energy under Contract No.
DE-AC03-76SF00098. We thank all these institutions for the generous allocation
of computer time.

\end{acknowledgements}

\end{document}